\documentclass[reprint,amsmath,amssymb,aps,prb,superscriptaddress,showpacs,citeautoscript,floatfix]{revtex4-1}
\usepackage[utf8]{inputenc}
\usepackage[ngerman, english]{babel}
\usepackage{graphicx}
\usepackage{dcolumn}
\usepackage{bm}
\usepackage{amsmath}
\usepackage{amssymb}
\usepackage[]{units}
\usepackage{subcaption}
\usepackage{booktabs}
\usepackage{lineno}
\usepackage{hyperref}
\graphicspath{
	{figures/}
}

\begin{document}


\title{Development and Evaluation of Interactive, Research-oriented Teaching Elements for Raising the Students' Interest in Research and for Facilitating the Achievement of Educational Objectives within the Lecture ``Atomistic Materials Modeling''\\[3ex] Entwicklung und Evaluation interaktiver, forschungsbezogener Lehrelemente zur Förderung des studentischen Forschungsinteresses und der Lernzielerreichung in der Vorlesung ''Atomare Materialmodellierung'' }


\author{Gregor Feldbauer$^{\,*,\,}$\href{https://orcid.org/0000-0002-9327-0450}{$^{\includegraphics[scale=0.5]{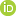}}$}$^{\,,\,}$}
\affiliation{Institute for Advanced Ceramics, TU Hamburg, Denickestraße 15, 21073 Hamburg, Germany}
\author{Marcel Steffen}
\affiliation{Institute for Transport Planning and Logistics, TU Hamburg}

\selectlanguage{english}
\begin{abstract}

The lecture ``Atomistic Materials Modeling'' is a core qualification of the master program ``materials science'' at the Hamburg University of Technology (TUHH). This lecture should be attended in the first semester of the program. About 10 students participated in the winter term 2016. Within the lecture, various modern methods for atomistic materials modeling are presented as well as their applications and limitations. Originally, the course was conceived as a traditional lecture. That didactic-methodical conception, however, does not seem to be ideal to support the students in reaching the educational objectives and to foster the students' interest in the covered topics as well as related research activities. A new didactic concept based on interactive engagement is designed to allow for a more individual and research-oriented learning experience of the students. To this end, team-work units involving worksheets and computer exercises are established replacing traditional lectures.  Additionally, the students get the possibility to sketch research proposals in small teams during their individual study time. By presenting their ideas, the students are enabled to gain bonus points for the final exam. The students are supposed to apply, discuss, and immerse themselves in selected topics of the lecture via those new elements. The effects of these innovations on the students' ability to reach the educational objectives and on their level of interest in related research activities are investigated in this work. Several questionnaires, observations by the lectures, and the results of the final exams are used as data sources allowing for a holistic examination. The analysis shows that the students are highly in favor of the new, interactive elements. Those elements support the students in reaching important educational objectives such as understanding of specific contents of the lecture. Moreover, the interest in research is increased. The questionnaires and exams, however, indicate some room for improvement. For example, the assessment of the limitations of different methods is difficult for the students. Consequently, an updated version of the presented concept including the findings of this work is supposed to be implemented in the future.

For your information, the language of this paper is German. 
\selectlanguage{ngerman} 

\begin{center}
\large{Kurzfassung}
\end{center}

Die Vorlesung ''Atomare Materialmodellierung'' ist eine Kernqualifikation im ersten Semester des Masterstudienganges ''Materialwissenschaften'' an der Technischen Universität Hamburg (TUHH) und wurde im Wintersemester 2016 von circa 10 Studierenden besucht. In dieser Vorlesung sollen die Studierenden moderne Methoden der atomistischen Materialmodellierung sowie deren Anwendung und Limitierungen kennenlernen. Ursprünglich wurde die Lehrveranstaltung als klassische Vorlesung konzipiert. Allerdings erscheint diese didaktisch-methodische Gestaltung nicht ideal, um die Lernziele zu erreichen und das Interesse der Studierenden an den behandelten Themen sowie entspre{\-}chender Forschung zu wecken. Ein neues didaktisches Konzept soll den Studierenden in interaktiven Elementen die selbstständige und forschungsnähere Beschäftigung mit der Thematik ermöglichen. Dazu werden während der Präsenzzeit Gruppenarbeiten mit Arbeitsblättern und Computerübungen eingeführt. Außerdem wird den Studierenden die Möglichkeit geboten, in der Eigenstudienzeit in Gruppen freiwillig einen Forschungsantrag zu skizzieren und durch dessen Präsentation \mbox{Bonuspunkte} zu erhalten. Durch diese Elemente sollen die Studierenden ausgewählte Inhalte der Vorlesung vertiefen, anwenden und diskutieren. Ob diese Neuerungen den Studierenden beim Erreichen der Lernziele helfen und das Interesse der Studierenden an Forschungstätigkeiten erhöhen, wird im Rahmen dieser Arbeit untersucht. Um eine ganzheitliche Betrachtung zu ermöglichen, werden mehrere Fragebögen, Beobachtung durch die Lehrenden sowie die benoteten Leistungen als Datenquellen herangezogen. Die Auswertung ergibt, dass die interaktiven Elemente von den Studierenden sehr positiv gesehen werden. Sie helfen den Studierenden beim Erreichen wichtiger Lernziele wie zum Beispiel dem Verständnis der Lehrinhalte. Außerdem erhöhen die neuen Elemente das Forschungsinteresse. Die Klausuren sowie Fragebögen zeigen allerdings auch weiteres Verbesserungspotential. Zum Beispiel fällt den Studierenden die Einschätzung der Grenzen der behandelten Methoden noch schwer. Das dargestellte Konzept soll in überarbeiteter Form in Zukunft weiterhin angewandt werden. 

\end{abstract}

\maketitle

\section{Einleitung und Ausgangssituation}\label{sec:intro}

In diesem Projekt, das als Bestandteil des Quali{\-}fizierungsprogramms ''Forschendes Lernen an der TUHH'' (Technische Universität Hamburg) durchgeführt wird, soll der Einfluss von interaktiven, forschungsbezogenen Elementen im Rahmen einer Vorlesung auf das Erreichen der Lernziele sowie auf das Forschungsinteresse der Studierenden untersucht werden. Als konkreter Untersuchungsgegenstand dient hier die Vorlesung ''Atomare Materialmodellierung''. Dies ist eine Kernqualifikation des im Wintersemester (WS) 2015/16 an der TUHH neu eingeführten Master-Studiengangs ''Materialwissenschaft'' des Maschinenbau Dekanats. Die Vorlesung ist dabei Teil des Moduls ''Materialphysik und atomare Materialmodellierung'', das zusätzlich noch aus der Vorlesung ''Materialphysik'' besteht. Das Modul wird mit einer gemeinsamen Klausur abgeschlossen. Die Absolvierung dieses thematisch physiknahen Pflichtmoduls mit einem Leistungsumfang von je drei ECTS pro Vorlesung ist für das erste Semester des Masterprogramms empfohlen. 

Die Lehrveranstaltung wird im WS 2016/17, das als Untersuchungszeitraum dient, von acht Studierenden regelmäßig besucht. Sie ist didaktisch als klassische Vorlesung konzipiert. Das Ziel der Vorlesung ist, dass die Studierenden die Konzepte verschiedener moderner Methoden der atomaren Materialmodellierung verstehen und deren Anwendung sowie Grenzen einschätzen können. Außerdem sollen die Studierenden fortgeschrittene Berechnungen von Eigenschaften atomarer Materialsysteme durchführen sowie geeignete Methoden dafür auswählen können. Es wird angenommen, dass eine reine Vorlesung das Erreichen dieser Lernziele und auch das Interesse der Studierenden an der Thematik nicht ausrei{\-}chend fördert. Daher wird in dieser Arbeit der Einfluss eines neuen didaktisch-methodischen Konzeptes, das interaktive Elemente in der Präsenzzeit und ein freiwilliges Projekt in der Eigenstudienzeit inkludiert, auf die zuvor genannten Aspekte untersucht. Diese neuen Elemente sind dabei alle als Gruppenarbeiten konzipiert, bei denen die Studierenden selbstständig unter minimaler Anleitung der Lehrenden Aufgabenstellungen bearbeiten. Das oben skizzierte Lehrproblem ist eine Annahme, die auf die Lehrerfahrung der Autoren sowie vorhandener Literatur (siehe Kapitel~\ref{sec:problem}) aufbaut, aber aufgrund mangelnder Evaluation früherer Jahrgänge auf keine Daten gestützt werden kann. Die Vorlesung wird in dieser Mo{\-}dulform erst zum zweiten Mal gehalten.

\section{Lehrproblem und Lösungsansatz}\label{sec:problem}

Besonders für das Erreichen der Lernziele, Berech{\-}nungen durchführen sowie geeignete Methoden auswählen zu können, scheint eine reine Vorlesung nicht die beste Form zu sein. Für beides sollte eine eingehende Auseinandersetzung mit den Methoden nötig und eine konkrete Anwendung dieser zumindest hilfreich sein. Auch die Begeisterung der Studierenden für die Vorlesungsthemen könnte darunter leiden, dass Methoden im Rahmen einer Vorlesung nur relativ abstrakt vorgestellt werden können. Vor allem kann so den Studierenden oft nicht vermittelt werden, wie eine Anwendung der Methoden konkret umgesetzt wird. Hier stellt sich die Frage, wie eine bessere Vermittlung statt{\-}finden könnte. Zur Auseinandersetzung mit dieser Frage wird auf Erkenntnisse der Lehrforschung zurückgegriffen. Eine Einführung zum Thema ''Forschung an der Lehre in der Physik'' wird zum Beispiel von  Robert Beichner gegeben.~\cite{beichner09} Die Hauptziele der Veränderungen an der Vorlesung sind die Ermöglichung eines nachhaltigeren Lernens und das verstärkte Wecken des Interesses der Studierenden. Ludwig Huber spricht in diesem Zusammenhang auch vom ''tiefen Lernen'', dem Aufbau von Wissen und Fähigkeiten, die nicht schnell vergessen werden, sondern dauerhaft abrufbar sind und flexibel auf neue Situationen angewandt werden können.~\cite{huber09} Dazu ist es typischerweise nötig, dass die Studierenden sich dieses Wissen selbst erarbeiten, organisieren sowie kritisch reflektieren. Zum Erreichen dieser Ziele bieten Methoden der forschungsbezogenen Lehre, auf diesen Begriff wird im Folgenden noch eingegangen, hervorragende Möglichkeiten.~\cite{schneider09} In einer groß angelegten Studie konnte Richard Hake zeigen, dass Studierende in interaktiven Lehrveranstaltungen deutlich besser abschnitten als Studierende in traditionellen Vorlesungen.~\cite{hake98} In Bezug auf die zu untersuchende Lehrveranstaltung wird folglich angenommen, dass die Studierenden bei der konkreten, forschungsnahen Anwendung verschiedener Methoden in interaktiven Elementen ein besseres Verständnis dieser Methoden und vor allem der einhergehenden Schwierigkeiten sowie Li-mitierung erreichen können. Außerdem sollen Hürden, die durch eine rein theoretische, teils abstrakte Beschreibung der Methoden entstehen können, abgebaut werden. Dies soll den Studierenden einen tieferen Einstieg in die Thematik, auch über die Vorlesung hinaus, erleichtern. 

\begin{figure*}
  \centering
    \includegraphics[width=0.98\linewidth]{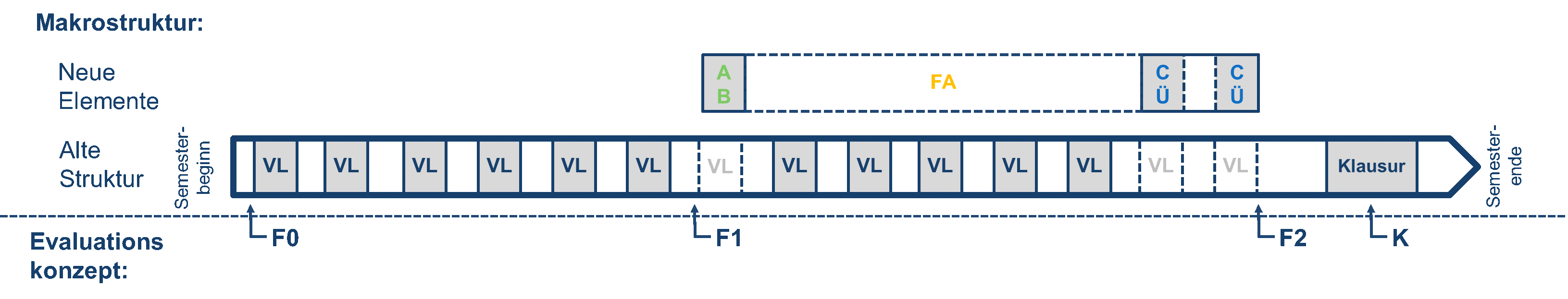}
  \caption{Änderung der Makrostruktur der Lehrveranstaltung durch Einführung der neuen, interaktiven Elemente: Arbeitsblätter (AB), Computerübungen (CÜ) und Forschungsantrag (FA). Die Zeitpunkte der Evaluation mittels Fragebögen (F0 -- F2) sind zusätzlich markiert. Als weitere Evaluationsmethoden dienen laufende Beobachtungen der Lehrenden sowie die abschließende Auswertung der Klausuren.}
  \label{fig:makroplan}
\end{figure*}

Auf diesen Ideen aufbauend wurde ein didaktisch-methodisches Konzept für die Vorlesung erarbeitet, das interaktive, forschungsbezogene Elemente enthält. Dazu werden drei Vorlesungseinheiten \`{a} 90 min sowie ein Teil der Eigenstudienzeit (etwa 10~\%) verwendet. Die Eigenstudienzeit war bisher nicht strukturiert und diente aus-schließlich der persönlichen Vorbereitung auf die Klausur. Ein Vergleich der alten und neuen Makrostruktur der Lehrveranstaltung ist in Abbildung~\ref{fig:makroplan} dargestellt. In der ersten interaktiven Einheit, die zur Hälfte des Semesters stattfindet, vertiefen die Studierenden in Vierergruppen anhand von Arbeitsblättern ihr Wissen zu einer in der Vorlesung behandelten Methode. Die Arbeitsblätter werden selbstständig unter minimaler Anleitung der Lehrenden bearbeitet. Informationen zum Konzept der minimalen Anleitung in der Lehre können zum Beispiel in Ref.~\onlinecite{aebli11} gefunden werden. Die Lehrenden treten in diesem interaktiven Element in enger Anlehnung an Friedewold \textit{et al.}~\cite{friedewold15b} als Lernbegleiter, d.h. als Rahmengestalter, Fachexperten, teilnehmende Zuhörer,  Diskussionspartner und Denkanstoßgeber, anstatt als klassische, vortragende Wissensvermittler auf. In dieser ersten Einheit geht es sowohl um die Erarbeitung und Vertiefung wichtiger theoretischer Grundlagen als auch um mögliche Anwendung-en dieser. Auf das so erarbeitete Grundlagenwissen bauen die weiteren interaktiven Elemente auf. 

In der zweiten und dritten interaktiven Einheit, die geblockt am Ende des Semesters durchgeführt werden, wenden die Studierenden in Zweiergruppen die in der ersten Einheit bearbeitete, computerbasierte Methode an konkreten, einfachen Beispielen an. Diese Einheiten haben Tutoriumscharakter und erlauben den Studierenden die Anwendung der Methode kennenzulernen und sich zwanglos damit auseinanderzusetzen. Dabei erfüllen die Studierenden die gestellten Aufgaben wiederum großteils selbstständig und die Lehrenden agieren nur unterstützend. Da aufgrund der zeitlichen Rahmenbedingungen in den bisher genannten interaktiven Elementen nur eine der vorlesungsrelevanten Methoden behandelt wird, ist zu erwarten, dass diese Elemente den Studierenden in erster Linie beim Verständnis dieser einen Metho{\-}de helfen können. Positive Effekte sind daher vor allem bezüglich dieser spezifischen Methode zu erwarten. 

Parallel zu den interaktiven Einheiten können die Studierenden in Kleingruppen von bis zu vier Personen freiwillig in der Eigenstudienzeit einen kurzen Forschungsantrag zu einem selbstgewählten Thema, das mit in der Vorlesung behandelten Methoden erforscht werden kann, erarbeiten. Das Ziel ist hier kein vollständig ausgearbeiteter Antrag, sondern eine Skizze. Dabei sollen die Studierenden ihr Wissen über die entsprechenden Methoden vertiefen und wichtige Aspekte des Forschungsprozesses allgemein kennenlernen. Zu diesen Aspekten zählen zum Beispiel die Umsetzung einer Idee in einen Forschungsplan, die dazugehörende Literaturrecherche, die Auswahl der passenden Metho{\-}den sowie die zeitliche und finanzielle Aufwandsabschätzung. Darüber hinaus wird die Teamfähigkeit im wissenschaftlichen Kontext trainiert. Zur Unterstützung werden die Studierenden bei Bedarf mit entsprechender Literatur versorgt, die als Startpunkt für ihre eigene Recherche dienen soll. Außerdem werden ihnen hilfrei{\-}che Werkzeuge wie das Planungsfünfeck~\cite{rienecker13}, ein Modell, das bei der Planung und Strukturierung eines Projektes hilft, gezeigt. Zusätzlich wird eine kurze Einführung in die Thematik der Forschungsanträge gegeben. Als Anreiz können die Studierenden durch die Präsentation ihrer Forschungsanträge Bonuspunkte für die Klausur sammeln. Die Präsentationen werden von den Lehrenden anhand vorab bekannt gegebener Kriterien beurteilt und entsprechend Bonuspunkte vergeben, die maximal 10~\% der bei der Klausur erreichbaren Punkte ausmachen.

Die Anwendung forschungsbezogener Lehre soll die Studierenden unterstützen, die Lernziele zu erreichen und sie darüber hinaus zu einer weiteren Beschäftigung mit vorlesungsrelevanten Themen und Forschungsaufgaben inspirieren.~\cite{Healey05} Mit forschungsbezogener Lehre ist hier Rueß~\textit{et al.} folgend Lehre mit einem Bezug zur Forschung gemeint, ohne eine Wertung über den Forschungsbezug zu beinhalten.~\cite{rues16} Laut der Klassifizierung nach Rueß~\textit{et al.} beinhalten bereits die reinen Vorlesungs{\-}einheiten der zu untersuchenden Lehrveranstaltung forschungsbezogene Elemente, bei der die Studierenden Forschungsergebnisse und -methoden rezeptiv kennenlernen. Durch die Einführung der interaktiven Elemente kommt die Anwendung der Ergebnisse und Methoden hinzu, sowie eingeschränkt auch die forschende Betätig{\-}ung mit den vorgestellten Methoden. Darüber hinaus erweitern die Einführung zu Forschungsanträgen und die Skizzierung eines solchen Antrages das Spektrum sowohl um die rezeptive und um die anwendende Aus{\-}einandersetzung mit dem Forschungsprozess als auch um die forschende Beschäftigung mit Forschungsergebnissen.  Bis auf die abschließende Ausführung des gesamten Forschungsprozesses werden damit die Teilbereiche der forschungsbezogenen Lehre nach Rueß~\textit{et al.} abgedeckt.~\cite{rues16} Dieses fehlende Element könnte zum Beispiel im Rahmen einer anschließenden Masterarbeit, die Ideen aus den Forschungsanträgen aufgreift, abgedeckt werden. Die Präsentation der Forschungsanträge überprüft außerdem das Erreichen des Lehrziels, dass die Studierenden in der Lage sein sollen, Ideen weiter zu entwickeln und Lösungen gegenüber Spezialisten präsentieren zu können. Im Rahmen der Präsentationen haben die Studierenden Gelegenheit, ihre Ideen zu diskutieren und Feedback sowohl von den Lehrenden als auch von ihren Kommilitonen zu erhalten. 

Für die Lehrenden stellt die Vorbereitung der interaktiven Elemente einen nicht zu unterschätzenden Aufwand dar. Vor allem die Ausarbeitung didaktisch wertvoller Aufgabenstellungen für die Arbeitsblätter und Computerübungen nimmt einige Zeit in Anspruch. Für diese Untersuchung konnte der zeitliche Aufwand in Grenzen gehalten werden, da auf bereits vorhandene Tutorien aufgebaut wurde. Diese Tutorien wurden im Rahmen des Graduiertenkollegs des SFB 986 von Wolfgang Heckel und Beatrix Elsner in Zusammenarbeit mit Alette Winter und Christian Kautz~\cite{herzog17} vom Zentrum für Lehre und Lernen an der TUHH erstellt und mussten für die untersuchte Vorlesung nur geringfügig angepasst werden. Die ursprüngliche Ausarbeitung der Unterlagen schlug allerdings mit mehreren Wochen zu Buche. Zur Betreuung der Studierenden ist  im Prinzip während der interaktiven Einheiten eine Lehrperson ausreichend. Aller{\-}dings sollte idealerweise eine Person pro ein bis zwei Klein{\-}gruppen anwesend sein, um einen vertiefenden, passgenauen Austausch sowie fortführende Diskussionen zu erleichtern. Der Aufwand für die Vorbereitung zu den Forschungsanträgen ist dagegen eher klein. Die Erstellung der Unterlagen sowie eine Literaturrecherche waren in wenigen Stunden erledigt. Der gesamte zusätzliche Aufwand für die Studierenden hält sich dagegen in Grenzen, da die interaktiven Einheiten während der Präsenzzeit stattfinden. Für den Forschungsantrag wird mit jeweils ungefähr einem Arbeitstag (etwa 10~\% der Eigenstudienzeit) pro Studierenden gerechnet. Diese eher knapp bemessene Zeit soll die Studierenden zur Verteilung der Aufgaben innerhalb der Gruppen animieren.

\section{Untersuchungsfrage und Hypothesen}\label{sec:hypo}

Wie oben beschrieben ist eine Leitfrage dieser Arbeit, welche Lehrmethoden eingesetzt werden können, um die Studierenden bei der Erreichung der Lernziele zu unterstützen - neben der Förderung des Forschungsinteresses. Nach dem Constructive-Alignment-Ansatz müssen dafür die Lehrinhalte, Lernziele und Prüfungen aufeinander abgestimmt sein.~\cite{biggs03} Dadurch kann ein Umfeld geschaffen werden, das Lernen fördert, um die gewünschten Lernergebnisse zu erzielen. Im konkreten Fall wird vor allem an den Lehrinhalten gearbeitet, indem interaktive, teils  forschungsnahe Elemente in die Vorlesung integriert werden und die Studierenden an einem Projekt (Forschungsantrag) arbeiten können. Die Präsentation{\-}en der Projekte bieten auch eine weitere Möglichkeit das Erreichen der Lernziele beurteilen zu können. Diese Veränderungen sind allerdings nur als erster Schritt auf dem Weg zu ''Constructive Alignment'' zu sehen. Für eine ganzheitliche Umsetzung sollte in weiteren Optimierungsrunden vor allem an der Überprüfung der Lernziele gearbeitet werden, da hier noch nicht das ganze Spektrum der Ziele adäquat abgedeckt wird. 

Die Untersuchungsfragen sind nun, ob die Einführung von interaktiven Teilen den Studierenden hilft, die Inhalte der Vorlesung zu verstehen sowie die Lernziele zu erreichen und ihr Interesse an der Thematik zu steigern. Dazu werden mehrere Hypothesen aufgestellt und anschließend untersucht. Die Hypothesen sind in drei Gruppen gegliedert. Teil 1 beschäftigt sich mit der Frage, ob die interaktiven Elemente beim Erreichen der Lernziele helfen. In Teil 2 geht es um das Interesse der Studierenden an Forschungstätigkeiten und den Einsatz von Lehrinhalten in der Forschung. Abschließend behandelt Teil 3 das Element Forschungsantrag explizit und die Akzeptanz der Umsetzung dieser Methode durch die Studierenden sowie die Eignung des  Elements im Rahmen forschungsbezogener Lehre.

\begin{itemize}
\item[] H1.1: Die interaktiven Elemente helfen den Studierenden die theoretischen Inhalte der Vorlesung zu verstehen. 

\item[] H1.2: Die interaktiven Elemente unterstützen die Studierenden die Grenzen der angewandten Metho{\-}de einschätzen zu können.

\item[] H2.1: Die interaktiven Elemente zeigen den Studierenden, wie Inhalte der Lehrveranstaltung in der Forschung angewandt werden.

\item[] H2.2: Die interaktiven Elemente erhöhen das Interesse der Studierenden an vorlesungsrelevanten Forschungsfragen.

\item[] H3.1: Die Studierenden nehmen das freiwillige Angebot zur Erarbeitung von Forschungsanträgen mehrheitlich an.

\item[] H3.2: Die Auseinandersetzung mit den Forschungsanträgen gibt den Studierenden Einblicke in Forschungsprozesse.
\end{itemize}

\section{Untersuchungsdesign}\label{sec:design}

Für diese Studie wird der Ansatz des Classroom Action Research (CAR) angewandt.~\cite{Mettetal12} Um eine ganzheitliche Untersuchung der Hypothesen zu ermöglichen, wird hierbei die Triangulation von Daten aus verschiedenen Untersuchungsmethoden verwendet.~\cite{denzin73, denzin12} Die Ergebnisse sind dabei allerdings nicht statistisch rigoros belastbar und beziehen sich in erster Linie auf die untersuchte Lehrveranstaltung. 

Einen Teil der zur Verfügung stehenden Daten bilden die bewertenden Leistungen (Klausur und Präsentation). Diese geben Aufschluss darüber, inwieweit die Studierenden die Lernziele erreicht haben und wo Probleme liegen. Die Ergebnisse sollen auch Rückschlüsse bezüglich Hypothese 1.1 erlauben. Da es sich bei den Lehrmethoden teils um neu eingeführte Werkzeuge handelt, wird durch die Beobachtung der Studierenden festgestellt, ob diese Methoden angenommen werden und insgesamt funktionieren. Dies beinhaltet zum Beispiel, ob sich die Studierenden aktiv beteiligen, ob Probleme auftreten und vor allem, ob und wie die Studierenden das Angebot einen Forschungsantrag zu erarbeiten annehmen. Diese Beobachtungen werden dabei nicht streng systematisch durchgeführt, sondern sollen lediglich den oben genannten Fragestellungen qualitativ Rechnung tragen und vor allem den Lehrenden ins Auge stechende Aspekte festhalten. Die Ergebnisse der Beobachtungen sollen ergänzend bei der Bewertung aller Hypothesen herangezogen werden. Als tragende Säule der Datenerhebung dienen Befragungen der Studierenden. Diese werden mit Fragebögen durchgeführt, die speziell für diese Studie konzipierten wurden. Dadurch wird mehrmals erhoben, ob die Studierenden die Lernziele erreichen, ob die neuen Methoden hilfreich sind und ob diese in der Lage sind, erhöhtes Interesse der Studierenden an den in der Vorlesung behandelten Themen zu wecken. Dabei werden sowohl die Meinungen und Selbsteinschätzungen der Studierenden als auch ihr Wissensstand abgefragt. Die mehrstufige Befragung, deren zeitlicher Verlauf in Abb.~\ref{fig:makroplan} skizziert ist, startet zu Beginn der Vorlesung. Mit dem ersten Fragebogen wird der fachliche Hintergrund und das Grundlagenwissen der Studierenden sowie ihr Interesse an der Vorlesungsthematik abgefragt. Eine zweite Befragung vor der ersten interaktiven Einheit soll klären, inwieweit die Vorlesung bis dahin die nötigen Inhalte zum Erreichen der Lernziele vermittelt und ob die Vorlesung das Interesse der Studierenden weckt oder sie eher abschreckt sich weiter mit der Thematik zu beschäftigen. In einer abschließenden Befragung am Ende des Semesters werden neuerlich das Erreichen der Lernziele und die Effekte der interaktiven Elemente überprüft. Außerdem haben die Studierenden die Möglichkeit, ihre Meinung zu den neuen Elementen zu äußern. Nach der abschließenden Triangulation der Erkenntnisse aus den verschiedenen Datenquellen werden die Hypothesen beurteilt und die durchgeführten interaktiven Elemente bezüglich ihrer Beibehaltung und Verbesserungsmöglichkeiten ergebnisoffen evaluiert.

Der zeitliche Aufwand für die Untersuchung der Neuerungen innerhalb der untersuchten Vorlesung summiert sich über das Semester verteilt auf einige Arbeitstage. Dabei schlagen vor allem die anfängliche Ausarbeitung des Konzepts und die Erstellung der drei Fragebögen, sowie deren Auswertung mit je{\-}weils mehreren Arbeitsstunden zu Buche. Die Verschriftlichung der Beobachtungen, die Analyse der bewerteten Leistungen sowie die Zusammenführung der Ergebnisse und eine abschließende Bewertung benötigen jeweils etwa eine Stunde. 

\section{Ergebnisse \& Diskussion}\label{sec:ergebnisse}

Im Folgenden werden die Ergebnisse der einzelnen Datenerhebungsmethoden präsentiert und kurz diskutiert. Anschließend werden die Ergebnisse zusammengeführt und gemeinsam behandelt. Für eine erste Bestandsaufnahme werden nun die Erkenntnisse der Beobachtungen durch die Lehrenden zusammengefasst. 

Während der Vorlesungseinheiten kam es laut dem Vortragenden zu wenig aktiver Teilnahme der Studierenden und teilweise zu geringen Teilnehmerzahlen. Besonders bei theorie- und/oder mathematiklastigen Abschnitten kam es auch zu verbalen und nonverbalen Unmutsäußerungen der Studierenden. Diese Beobachtung{\-}en deuten auf ein Lehrproblem hin und untermauern die ursprüngliche Annahme, dass die Konzeption der Lehrveranstaltung als reine Vorlesung nicht ideal ist. Während der interaktiven Einheiten ergaben sich dazu eher konträre Beobachtungen. Die Studierenden arbeite{\-}ten sehr fleißig  und motiviert an den gestellten Aufgaben. Es wurde in den Gruppen ausgiebig diskutiert und bei Unklarheiten wurden die Lehrenden um Hilfe gebeten. Zudem entwickelten sich interessante Fachgespräche über die unmittelbaren Aufgaben hinaus. In Bezug auf die Forschungsanträge zeigte sich ein ähnliches Bild. Die Studierenden, die in zwei Gruppe je einen Antrag erarbeitetet hatten, präsentierten sehr umfangreich ihre Ideen, bei denen sie nicht auf die Themenvorschläge der Lehrenden zurückgegriffen hatten. Da die Mehrheit der Studierenden sich an einem Forschungsantrag beteiligte, ist H3.1 damit bestätigt. Bei der Einschätzung der Machbarkeit und der Metho{\-}denwahl offenbarten sich allerdings einige Schwächen. Bei der Präsentation und der anschließenden Diskussion zeigten sich auch deutliche Unterschiede innerhalb der Arbeitsgruppen bezüglich der Auseinandersetzung mit dem vorgestellten Thema. Diese Gefahr bei Gruppenarbeiten sollte verstärkt in die zukünftige Planung einbezogen werden. Wie viel Zeit die Studierenden für das Element Forschungsantrag aufwendeten, wurde nicht erhoben. Insgesamt ergaben die Beobachtungen ein sehr positives Bild der neu eingeführten Elemente, wobei es weiteres Verbesserungspotential gibt. Bei der Vorlesung werden größere Herausforderungen sichtbar, an deren Lösung weitergearbeitet werden sollte.

\begin{figure}
  \centering
    \includegraphics[width=1.\linewidth]{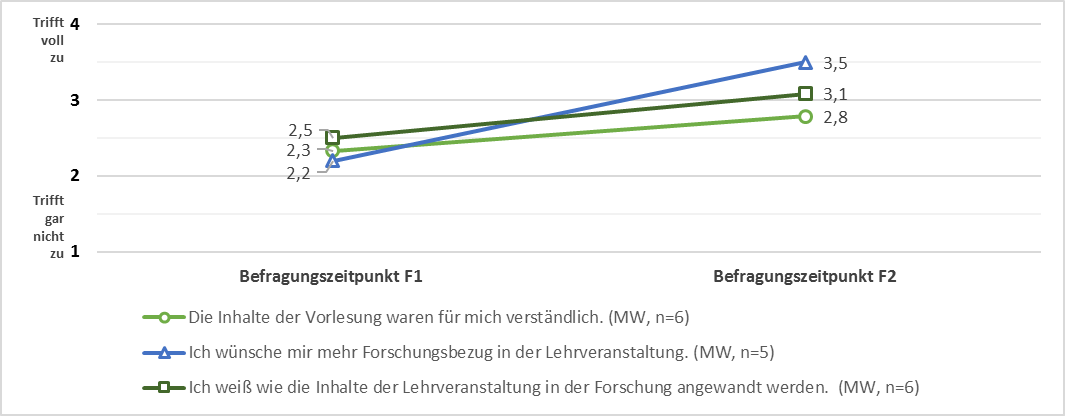}
  \caption{Mittelwerte der Antworten der Studierenden auf drei Fragen an den Evaluationspunkten F1 (N = 6) und F2 (N = 7). \textit{N} gibt die Zahl der Befragungsteilnehmenden an dem jeweiligen Evaluationspunkt an, während \textit{n} die Zahl der gültigen Antworten auf eine konkrete Frage ist.}
  \label{fig:diag1}
\end{figure}

\begin{figure}
  \centering
    \includegraphics[width=1.\linewidth]{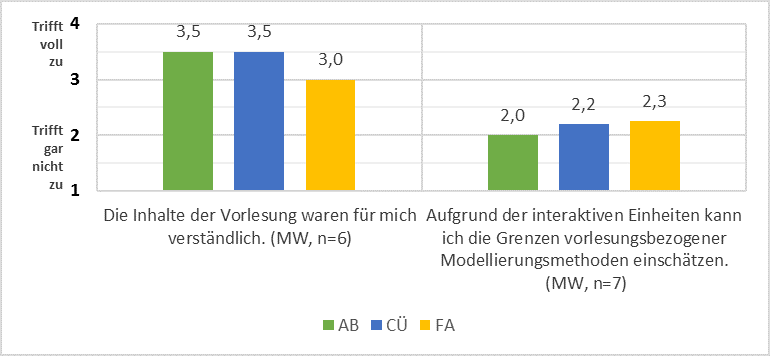}
  \caption{Mittelwerte der Antworten der Studierenden auf zwei Fragen, die den Hypothesenblock H1 betreffen, am Evaluationspunkt F2 (N = 7). \textit{N} gibt die Zahl der Befragungsteilnehmenden an dem jeweiligen Evaluationspunkt an, während \textit{n} die Zahl der gültigen Antworten auf eine konkrete Frage ist.}
  \label{fig:diag2}
\end{figure}

\begin{figure}
  \centering
    \includegraphics[width=1.\linewidth]{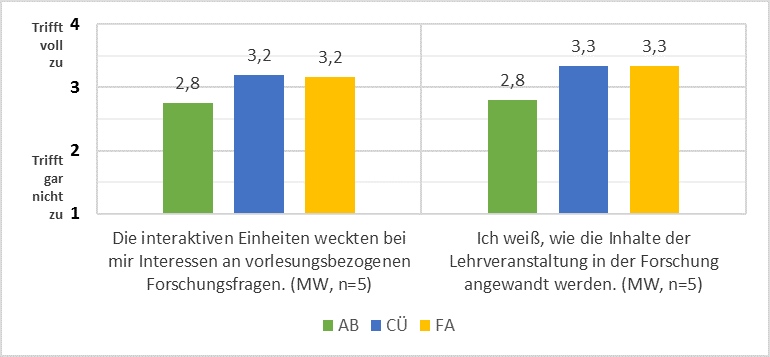}
  \caption{Mittelwerte der Antworten der Studierenden auf zwei Fragen, die den Hypothesenblock H2 betreffen, am Evaluationspunkt F2 (N = 7 ). \textit{N} gibt die Zahl der Befragungsteilnehmenden an dem jeweiligen Evaluationspunkt an, während \textit{n} die Zahl der gültigen Antworten auf eine konkrete Frage ist.}
  \label{fig:diag3}
\end{figure}

\begin{figure}
  \centering
    \includegraphics[width=1.\linewidth]{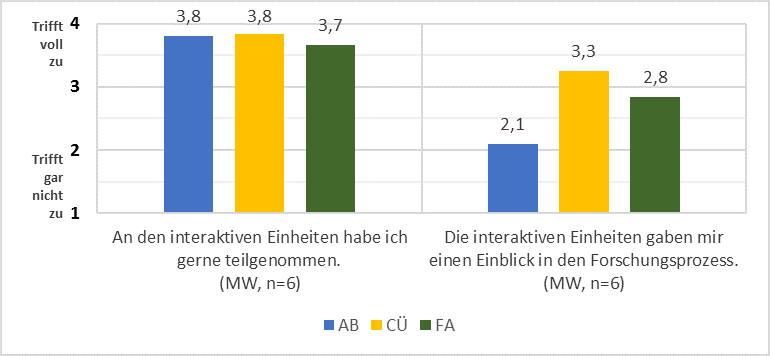}
  \caption{Mittelwerte der Antworten der Studierenden auf zwei Fragen, die den Hypothesenblock H3 betreffen, am Evaluationspunkt F2 (N = 7). \textit{N} gibt die Zahl der Befragungsteilnehmenden an dem jeweiligen Evaluationspunkt an, während \textit{n} die Zahl der gültigen Antworten auf eine konkrete Frage ist.}
  \label{fig:diag4}
\end{figure}

Der erste Fragebogen zu Vorlesungsbeginn zeigte die Heterogenität der Gruppe bezüglich ihres Wissensstands. Zu den vertretenen Bachelorabschlüssen gehörten z.B. Chemie, Maschinenbau, Material- und Nanowissenschaften. Die Studierenden wiesen gutes Wissen im Bereich der Materialwissenschaft auf, wenig aber dagegen in den Bereichen der Materialmodellierung und Quantenphysik. Hier deckten sich die Selbsteinschätzungen der Studierenden mit den Ergebnissen der Wissensfragen. Im Bereich der Festkörperphysik überschätzten sich die Studierenden leicht, zeigten aber Vorwissen. Die Befragung ergab ein prinzipielles Interesse an den Inhalten der Vorlesung, wobei die konkreten Erwartung{\-}en weit auseinandergingen. Die zweite Befragung vor der ersten interaktiven Einheit bestätigte das prinzipielle Interesse der Studierenden an der Vorlesungsthematik und Forschungstätigkeiten allgemein. Die Bedeutung der Lehrinhalte war den Studierenden eher klar, allerdings blieben konkrete Umsetzungen eher unklar und teils fehlte das Verständnis der Inhalte. Bei den offenen Fragen zeigte sich erneut die Heterogenität der Gruppe. So waren für einige das Tempo und die Inhaltsdichte der Vorlesung wegen fehlenden Vorwissens zu hoch, während andere sich einen tieferen Einstieg in die Thematik wünschten. Insgesamt wurde  mehr Bezug zu Anwendungen gewünscht und die Einführung von Übungen als sinnvoll erachtet. Der dritte Fragebogen nach den interaktiven Einheiten am Ende der Lehrveranstaltung sollte vor allem der Einschätzung der interaktiven Elemente dienen. Eine grafische Zusammenfassung der wichtigsten Ergebnisse wird in den Abb.~\ref{fig:diag1} bis ~\ref{fig:diag4} gegeben. Insgesamt nahmen die Studierenden sehr gerne an diesen Elementen teil (Abb.~\ref{fig:diag4}). Im Folgenden werden für die interaktiven Elemente Abkürzungen verwendet: Arbeitsblätter (AB), Computerübung{\-}en (CÜ) und Forschungsantrag (FA). Positiv gaben die Studierenden folgende Punkte an: Die Elemente halfen beim Verständnis der Inhalte (Abb.~\ref{fig:diag1} und laut Abb.~\ref{fig:diag2} vor allem AB und CÜ). Das Wissen zu Anwendungen stieg (Abb.~\ref{fig:diag1} und laut Abb.~\ref{fig:diag3} vor allem durch CÜ und FA). Das Interesse an Forschungsfragen wurde geweckt (laut Abb.~\ref{fig:diag3} vor allem durch CÜ und FA) und Einblicke in Forschungsprozesse wurden erlaubt (laut Abb.~\ref{fig:diag4} vor allem durch CÜ und FA). Allerdings schafften es die Elemente zum Beispiel nicht, die Grenzen der Metho{\-}den adäquat zu vermitteln (siehe Abb.~\ref{fig:diag2}). Von den abgefragten Lernzielen war dies das Einzige, bei dessen Erfüllung keines der interaktiven Elemente mehrheitlich half. Hier besteht eindeutig Verbesserungsbedarf. Laut Abb.~\ref{fig:diag1} stieg der Wunsch nach mehr Forschungsbezug im Zeitraum der interaktiven Einheiten. Der Grund dafür kann hier nicht empirisch angeben werden. Allerdings ist eine mögliche Interpretation, dass die interaktiven Elemente Möglichkeiten aufzeigten und so den Studierenden Lust auf mehr Forschungsbezug machten. Bei den offenen Fragen bestätigte sich das bisherige Bild. Die interaktiven Elemente wurden sehr positiv aufgenommen und einige Verbesserungsmöglichkeiten aufgezeigt. Beispiele hierfür sind, dass die Aufgaben für die Gruppenarbeiten bereits vorab zugänglich gemacht werden sollten und diese an manchen Stellen verbessert werden könnten. Außerdem wurde mehrmals mehr Zeit für die interaktiven Elemente, hier vor allem für die CÜ sowie den FA, gewünscht. Neuerlich wurde vor allem fehlendes Vorwissen als Hindernis für das Verstehen aller Lehrinhalte gesehen und mehr Zeit für weitere Übungen gewünscht.

Als dritte Datenquelle wurden die Ergebnisse der Klausuren zum Modulteil ''Atomare Materialmodellierung''  ausgewertet. Die Studierenden, die an interaktiven Einheiten teilgenommen hatten, erhielten Noten zwischen 2,3 und 3,3. Dies ist prinzipiell ein gutes Ergebnis, allerdings bleibt noch Luft nach oben und vor allem das Ausbleiben von positiven Ausreißern ist auffällig. Zu denken gibt die Tatsache, dass sich durchgehend Proble{\-}me mit einem Themenbereich zeigten. Interessant ist, dass dieser Komplex sowohl in der Vorlesung als auch in der Klausur an letzter Stelle kam. Eine genauere Analyse dieses Umstandes ist mit den vorhandenen Daten aller{\-}dings nicht möglich. Allgemein zeigten die Ergebnisse, dass es vor allem Schwierigkeiten mit Aufgaben gab, die spezifische Fragen zu Themen jenseits der klass{\-}ischen Physik sowie zu spezifischen theoretischen Grundlagen verschiedener Modellierungsmethoden darstellten. Die Fragebögen legen nahe, dass dies an fehlendem Vorwissen liegen könnte und in der Vorlesung dies aus zeitlichen Gründen nicht ausreichend nachgeholt werden konnte. Außerdem zeigen sich Probleme beim Verständnis einiger Lehrinhalte. Hier sticht eine Aufgabe posi{\-}tiv heraus, deren Beantwortung durch die Bearbeitung der Arbeitsblätter in der ersten interaktiven Einheit erleich{\-}tert wurde. Bei dieser Frage erreichten die Studierenden durchschnittlich etwa zwei Drittel der möglichen Punkte, während bei ähnlichen Aufgaben dieser Schnitt auf meist deutlich unter die Hälfte sank. Dies deutet auf den posi{\-}tiven Einfluss der interaktiven Einheiten hin und darauf, dass diese beim Verständnis der Inhalte helfen. Der Hinweis gilt hier natürlich direkt nur für die Arbeitsblätter. Da aufgrund der zeitlichen Rahmenbeding{\-}ungen in den interaktiven Elementen Arbeitsblätter und Computerübung nur eine der vorlesungsrelevanten Metho{\-}den behandelt wurde, können diese Elemente bezüglich des Verständnisses der anderen Methoden keinen großen Beitrag lei{\-}sten.  

Insgesamt kann die Einführung der interaktiven Einheiten positiv bewertet werden. Sie stießen bei den Studierenden auf große Zustimmung (siehe Abb.~\ref{fig:diag4}) und halfen beim Erreichen wichtiger Lernziele wie dem Verständnis der Lehrveranstaltungsinhalte. Dies lässt sich durch die Aussagen der Studierenden in den Fragebögen sowie den Beobachtungen und Klausurauswertungen gut untermauern. Laut der Studierenden wurden die Vorlesungsinhalte im Zeitraum, in dem die interaktiven Einheiten stattfanden, verständlicher (siehe Abb.~\ref{fig:diag1}) und diese Einheiten unterstützten explizit diese Erhöhung des Verständnisses (siehe Abb.~\ref{fig:diag2}). H1.1 kann somit bestätigt werden.  Darüber hinaus gaben die Studierenden an, dass die interaktiven Elemente, hier vor allem CÜ und FA, ihnen vermittelten, wie Inhalte der Lehrveranstaltung in der Forschung angewandt werden. Wie im vorigen Punkt stiegen die Zustimmungswerte im Zeitraum der interaktiven Einheiten und deren positiver Einfluss darauf wurde explizit angegeben (siehe Abb.~\ref{fig:diag1} und~\ref{fig:diag3}). H2.1 kann daher bestätigt werden. Außerdem erhöhten die interaktiven Elemente, wiederum speziell CÜ und FA, das Interesse der Studierenden an vorlesungsbezogenen Themen und Forschungsfragen (siehe Abb.~\ref{fig:diag3}). Somit ist auch H2.2 bestätigt. Die Tatsache, dass die Arbeitsblätter hier nicht so positiv wie die anderen Elemente beitragen, ist nicht überraschend, da die AB vor allem Grundlagen vertiefen sollen. Laut der Antworten aus den Fragebögen gaben die interaktiven Elemente CÜ und FA den Studierenden auch Einblicke in den Forschungsprozess, während AB dies, wie geplant, nicht leistete (siehe Abb.~\ref{fig:diag4}). Folglich ist H3.2 bestätigt. Hier soll noch angemerkt sein, dass CÜ und FA unterschiedliche Einblicke in den Forschungsprozess geben. Während die CÜ die Anwendung einer Forschungsmethode vermittelt, zeigt der FA Abläufe des Forschungsprozesses an sich auf. Der schlechtere Zustimmungswert bei FA gegenüber CÜ deutet hier auf Optimierungsmöglichkeiten in der Lehrplanung hin. Allerdings zeigte sich auch deutlicheres Verbesserungspotential bei den interaktiven Elementen, um das Erreichen weiterer Lernziele wie die Einschätzung der Grenzen verschiedener Methoden zu unterstützen (siehe Abb.~\ref{fig:diag2}). Das Verfehlen dieses Lernzieles zeigte sich auch wie oben erwähnt in den Ergebnissen der Forschungsanträge. H1.2 kann daher für die aktuelle Form der Lehrveranstaltung nicht bestätigt werden. Die Lehrveranstaltung insgesamt muss kritischer bewertet werden. Die Studierenden haben teils Probleme, den Inhalten zu folgen. Dies kann teilweise durch fehlendes Vorwissen erklärt werden. Aufgrund der Heterogenität der Gruppe, die durch die Auslegung des Masterprogramms vorprogrammiert ist, lässt sich eine Lösung nicht ohne weiteres finden.

\section{Zusammenfassung, Fazit \& Ausblick}\label{sec:zsfg}

Im Rahmen dieses Projektes wurden das didaktisch-methodische Konzept der Vorlesung ''Atomare Materialmodellierung'' überarbeitet und die Auswirkungen auf das Erreichen der Lernziele und das Interesse der Studierenden an vorlesungsbezogenen Themen und Forschung untersucht. Der ursprünglich reine Vorlesungscharakter der Lehrveranstaltung wurde durch neue interaktive, forschungsnahe Elemente aufgebrochen. Dazu wurden drei Vorlesungseinheiten durch Gruppenarbeiten mit Arbeitsblättern und Computerübungen ersetzt, um so die Hintergründe der Methoden zu vertiefen und konkret anzuwenden. Darüber hinaus wurde die Möglichkeit geschaffen, in der Eigenstudienzeit einen Forschungsantrag zu skizzieren und durch dessen Präsentation Bonuspunkte für die Klausur zu sammeln. Diese interaktiven Elemente sind vielseitig einsetzbar und es wird keine Einschränkung auf bestimmte Vorlesungen oder Fachrichtungen gesehen. 

Zur Evaluierung der Neuerungen und zur Untersuchung der Auswirkungen wurden Daten mit verschiedenen Methoden erhoben und anschließend ausge{\-}wertet. Als Datenquellen dienten von den Studierenden zu drei Zeitpunkten beantwortete Fragebögen, die Beobachtungen der Lehrenden und die Auswertung der Klausuren. Die Analyse der Daten und die Triangulation der Ergebnisse zeichnen im Großen und Ganzen ein positives Bild der neuen Elemente. Die Studierenden nahmen die interaktiven Elemente sehr positiv auf und beteiligten sich tatkräftig daran. Die Elemente konnten die Studierenden beim Erreichen wichtiger Lernziele unterstützen und ihr Interesse an vorlesungsrelevanten Themen sowie der Forschung fördern. Die Evaluation zeigte allerdings auch Verbesserungspotential sowohl bei den neuen Elementen als auch bei der Vorlesung allgemein. Vor allem konnte das Erreichen einiger Lernziele - wie zum Beispiel die Grenzen verschiedener Methoden - noch nicht ausreichend begünstigt werden. Obwohl die neuen Elemente den Forschungsbezug quantitativ als auch qualitativ erhöhen, sind weitere Verbesserungen in Zukunft möglich, um Forschungsaspekte ganzheitlicher in die Lehrveranstaltung zu integrieren. Hier wäre zum Beispiel denkbar, dass die Studierenden während des Semesters ein konkretes Forschungsprojekt bearbeiten oder die interaktiven Elemente ergebnisoffener und noch forschungsnäher gestaltet werden. Im Sinne einer nachhaltigen Entwicklung der Lehrveranstaltung sollen diese Ergebnisse in die Planung für das nächste Semester einfließen und entsprechende Adaptierungen umgesetzt werden. Zur Verbesserung der ganzen Lehrveranstaltung sind verschiedene Optionen denkbar, die einzeln oder gemeinsam umgesetzt werden könnten und explizit von einzelnen Studierenden angeregt wurden. Es könnte ein stärkerer Fokus auf Grundlagenwissen gelegt werden. Hier besteht natürlich die Gefahr, dass ein Teil der Studierenden unterfordert wird. Allerdings zeigte die Klausur hier durchgehende Defizite. Ohnehin ist die Wiederholung von Inhalten wichtig für ein erfolgreiches Lernen. Bei den Lernzielen würde dann das Verstehen der Inhalte noch mehr in den Vordergrund treten. Eine Ausweitung der interaktiven Elemente wäre eine weitere Option. So könnten mehr vorlesungsrelevante Methoden angewandt werden. Dies sollte auch bei der Einschätzung und  Abgrenzung der Methoden helfen. Eine ausgewei{\-}tete Präsentation konkreter Anwendungsbeispiele konnte bei diesen Aspekten ebenfalls unterstützend wirken. Die Umsetzung dieser Ideen wäre innerhalb des bestehenden Rahmens möglich, allerdings müssten einige Inhalte der Vorlesung gekürzt oder gestrichen werden um Raum für die Neuerungen zu schaffen. Außerdem wäre die Einführung von Übungen als zusätzliche, die aktuelle Vorlesung erweiternde Lehrveranstaltung eine denkbare Möglichkeit. Hierfür müsste allerdings teilweise der Lehrplan angepasst werden und der Lehrveranstaltung mehr Raum gegeben werden. Dies ist natürlich eine komplexe Aufgabe, die nur gemeinsam mit den Verantwortlichen für das Model und den Studiengang sowie den übergeordneten Universitätsorganen erreicht werden kann. Die vorgeschlagenen Anpassungen des Lehrplans werden aber als sehr sinnvoll erachtet, da ohne eine entsprechende Grundlagenausbildung eine weitere Vertiefung nur schwer möglich ist.

Im Hinblick auf die nachhaltige Entwicklung der Lehrveranstaltung sollten auch in den kommenden Jahren alte sowie neue Lehrelemente untersucht werden. So stellt sich zum Beispiel beim Element Forschungsantrag die Frage, wie die Studierenden die Aufgaben innerhalb der Gruppen verteilen und wie viel Zeit sie für dieses Projekt aufwenden. Dies sollte zeigen, ob die veranschlagte Zeit angemessen ist und warum die Studierenden anscheinend unterschiedlich von diesem Projekt in Bezug auf die Klausurleistung profi{\-}tierten. Außerdem könnten die Effekte der einzelnen interaktiven Elemente auf die Lernzielerreichung und das Forschungsinteresse noch gezielter überprüft werden, zum Beispiel durch Anpassungen am Evaluations{\-}konzept. Da solche Aspekte durch die aktuelle Studie nun bekannt sind, kann darauf in den folgenden Jahren spezifischer eingegangen werden. Dies kann zum Beispiel mit detaillierten Fragen im Evaluierungsprozess erreicht werden. Die Gültigkeit der in dieser Studie präsentierten Ergebnisse sollte aber auch in Zukunft für neue Gruppen oder Vorlesungen überprüft werden, da es vor allem in relativ  kleinen Gruppen zu deutlichen Unterschieden zwischen verschiedenen Jahrgängen kommen kann.

Abschließend kann festgehalten werden, dass die neuen interaktiven Elemente in der Lage sind, die Studierenden bei der Erreichung der Mehrheit der Lernziele zu unterstützen und ihr Forschungsinteresse zu wecken. Basierend auf den gewonnenen Erkenntnissen werden die Elemente in Zukunft weiter optimiert, damit die Studierenden noch ganzheitlicher gefördert werden.

\section*{Danksagung}\label{sec:acknow}
Besonderer Dank gebührt Frank Lechermann dafür, dass die Vorlesung ''Atomare Materialmodellierung'' als Untersuchungsgegenstand für diese Studie dienen konnte und für die gewährte Freiheit bei der Einführung von neuen Lehrelementen. Wir bedanken uns besonders bei den Studierenden, die an den interaktiven Ele{\-}menten teilgenommen haben und ohne deren Feedback und der Beantwortung der Fragebögen diese Studie in dieser Form nicht möglich gewesen wäre. Des Weiteren möchten sich die Autoren beim ZLL-Team der TUHH, im Besonderen bei Ulrike Bulmann, für die Betreuung im Rahmen des Qualifizierungsprogramms ''Forschendes Lernen an der TUHH'' bedanken. Außerdem danken wir Kerstin Rosenberger und Wiebke Rüther für wertvolles Feedback und Diskussionen zu dieser Arbeit. GF bedankt sich für die Unterstützung durch die Deutsche Forschungs Gesellschaft (DFG) im SFB~986~``M$^3$'', Projekt~A4, sowie bei Beatrix Elsner und Wolfgang Heckel für die Zurverfügungstellung der Tutorien sowie für die Mithilfe bei deren Bearbeitung und Durchführung. 

\section*{Informationen zu den Autoren}\label{sec:autoren}

\noindent{\textbf{Korrespondierender Autor}}\\
$^*$E-mail: \href{mailto:gregor.feldbauer@tuhh.de}{gregor.feldbauer@tuhh.de}\\
\textbf{ORCID}$^{\includegraphics[scale=0.5]{orcid_16x16.png}}$\\ 
Gregor Feldbauer: \href{https://orcid.org/0000-0002-9327-0450}{0000-0002-9327-0450}\\

\bibliographystyle{unsrt}
\bibliography{refs.bib}

\end{document}